\pdfoutput=1
\documentclass[10pt,a4paper]{article}
\usepackage{graphicx}
\usepackage{subfig}
\title{The Trade-offs with Space Time Cube Representation of Spatiotemporal
    Patterns}
\author{Per Ola Kristensson, Nils Dahlb\"ack, Daniel Anundi,\\
    Marius Bj\"ornstad, Hanna Gillberg, Jonas Haraldsson,\\
	Ingrid M\aa{}rtensson, Matttias Nordvall, Josefin St\aa{}hl\\
\\
Department of Computer and Information Science\\
Link\"{o}ping University\\
Link\"{o}ping, Sweden}
\begin{document}
\maketitle

\begin{abstract}
Space time cube representation is an information visualization technique where
spatiotemporal data points are mapped into a cube. Fast and correct analysis
of such information is important in for instance geospatial and social
visualization applications. Information visualization researchers have
previously argued that space time cube representation is beneficial in
revealing complex spatiotemporal patterns in a dataset to users. The argument
is based on the fact that both time and spatial information are displayed
simultaneously to users, an effect difficult to achieve in other
representations. However, to our knowledge the actual usefulness of space time
cube representation in conveying complex spatiotemporal patterns to users has
not been empirically validated. To fill this gap we report on a
between-subjects experiment comparing novice users’ error rates and response
times when answering a set of questions using either space time cube or a
baseline 2D representation. For some simple questions the error rates were
lower when using the baseline representation. For complex questions where the
participants needed an overall understanding of the spatiotemporal structure
of the dataset, the space time cube representation resulted in on average
twice as fast response times with no difference in error rates compared to the
baseline. These results provide an empirical foundation for the hypothesis
that space time cube representation benefits users when analyzing complex
spatiotemporal patterns.
\end{abstract}

\section{Introduction}
The space time cube is an information visualization technique that displays
spatiotemporal data inside a cube, sometimes called an ``aquarium''
\cite{hagerstrand:1975}. The
height axis is used to denote time. The space time cube was originally
proposed by Torsten H\"agerstrand in the early 1970s in a seminal paper on time
geography \cite{hagerstrand:1970}, and has since then been mainly used to display geospatial data
\cite{miller:2005}. The space time cube representation has been proposed by
Kraak \cite{kraak:2003} and
others \cite{gatalsky:2004,kapler:2004} as a tool in spatiotemporal visualization [3,4,5]. Recent
applications of space time cube representation include geospatial
visualization \cite{gatalsky:2004,kapler:2004,kraak:2003} and visualization of
sport \cite{moore:2003}. Figure \ref{fig:stc} shows an
example of space time cube representation.

Information visualization researchers have stated that the theoretical
advantage of the space time cube is the ability to efficiently convey complex
spatiotemporal patterns to users \cite{gatalsky:2004,kapler:2004,kraak:2003}. The argument is based on the fact
that space time cube representation presents users with the full
spatiotemporal dataset in a single view, in contrast to traditional 2D
displays where complex spatiotemporal information is often conveyed using time
slider controls, animation or resolution-limited pseudocolor sequences
\cite{ware:2004}.

However, before we research and build complex space time cube applications a
solid understanding of the costs and benefits of presenting users with space
time cube representation is desirable. To our knowledge no formal empirical
experiment comparing space time cube against a baseline 2D visualization has
been carried out. Hence we do not know if there is any advantage at all in
using space time cube representation. As has recently been argued in the
literature (e.g. \cite{wijk:2006}), evaluation is an important contribution
towards changing some parts of the information visualization field into a
``hard'' science.

\subsection{Contributions}
In this paper we present empirical results from a baseline comparison where we
investigate users’ ability to quickly and correctly answer a set of questions
in varying difficulty and complexity about a dataset in the continuous
spatiotemporal domain. 

We provide empirical data that highlight the trade-offs in space time cube
representation. Our results show that space time cube representation results
in more errors for novice users answering a category of “simple” questions
such as ``Are two persons at the same place at 9:00?'' More interestingly, the
results also reveal that using space time cube representation the average
response times were reduced from 121 s to 60 s when novice users were asked to
answer questions that required an overall understanding of the spatiotemporal
patterns in the dataset. The latter result supports the claim that space time
cube representation is advantageous in conveying complex spatiotemporal data
to users. Further, it motivates research and evaluation of new space time cube
representations for a plethora of application domains.

\subsection{Research Questions}
Given the lack of foundation from previous empirical research results, we
decided to focus this investigation on the most basic questions:

\begin{enumerate}
\item Can novice users understand and use a space time cube system effectively
after a short amount of practice?
\item Are there measurable performance differences in terms of error rates and
response times between a space time cube system and a baseline 2D system?
\item Are there measurable performance differences in terms of error rates and
response times between a space time cube system and a baseline 2D system for
specific categories of questions?
\end{enumerate}

The two dependent variables were error and response time. Error is indicative
of if users understood the dataset under a particular visualization. Response
time shows how long it took participants to make an informed decision
using a particular visualization. Relative longer response times indicate that a
particular visualization was not as efficient. Unlike error, the significance of response
time may differ for different tasks (as long as the response time difference is
not severe, e.g. one minute vs. one hour).

We decided to concentrate on novice users for the following reasons. First, it
is hard to find expert users that have proficiency in either a space time cube
system, or in another visualization system that can be used as a suitable
baseline \cite{tory:2004}. Most likely, expert users have varying knowledge of an ensemble
of different visualization systems and tools, creating difficulties for a
direct comparison between two systems. Second, if we can show that space time
cube representation is advantageous to novice users, such a result is in
itself useful as an empirical building block: researchers then know that
novice users understand space time cube representation relatively quickly and
can easily recruit non-expert participants for many different experimental
setups. Third, if novice users are shown to use space time cube representation
effectively, there is no reason to believe expert users would not be able to
do the same. In fact, expert users are most likely even better.

Note that we do not rule out the possibility of a study of expert users’
experience with space time cube representation. However, we do believe such a
study is probably more interesting from another perspective, for example, to
study \emph{how} expert users analyze complex spatiotemporal patterns.

\section{Domain}
\begin{figure}
\begin{center}
\includegraphics[width=10cm]{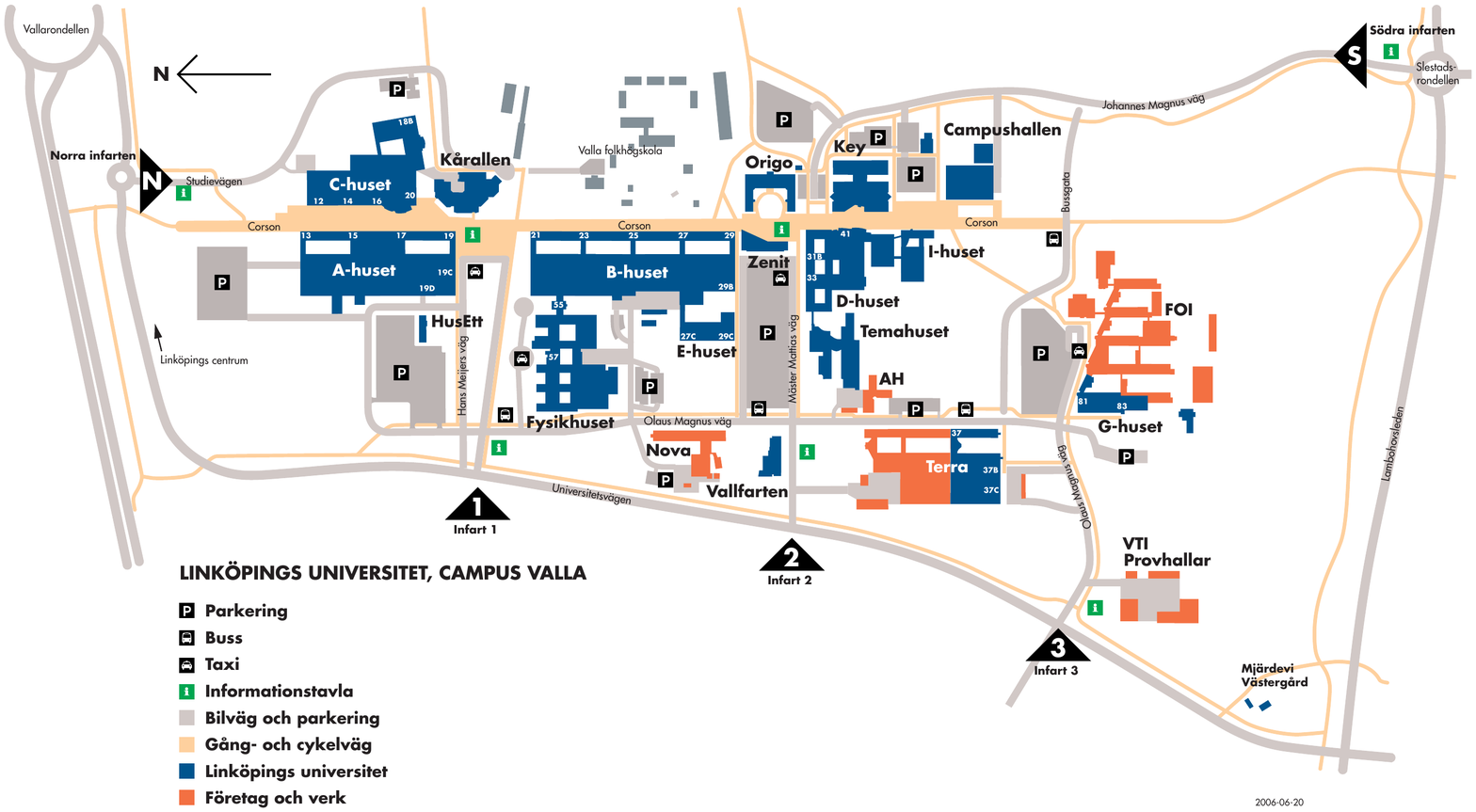}
\end{center}
\caption{The campus map used in the experiment. This campus map is the official
campus map, which is used by the university in brochures, on the web and on
notice boards.}
\label{fig:map}
\end{figure}

We decided to use human walking traces overlaid on a schematic of a university
campus area as our domain. Figure \ref{fig:map} shows the campus map. Note
that we cropped the outside areas of the map (e.g.~road entrances). Clearly,
the map choice may affect experimental results. Different maps can be designed
for many different purposes, and no map is ``perfect'' unless (possibly) it is
specifically tailored for a particular set of analytic questions. To avoid
this issue altogether we settled for using the official campus map that was
designed by university staff and has been in use on, for example, notice boards all
over the campus for many years. A walking data analysis application is 
realistic in practice given recent interest in social visualization. For
example, Aipperspach \cite{aipperspach:2006} describes recent work on
visualizing walking data. We acquired the walking data by tracking volunteer students' movement along
the campus during a day.

\section{Systems}
A dilemma when comparing a visualization method such as space time cube
representation with another is the choice of a suitable baseline. Clearly no
baseline representation will ever be ``fair'' from all perspectives of
information visualization. This dilemma makes costly empirical experiments
risky and may be a factor influencing the limited number of user studies in
the information visualization field \cite{tory:2004}. In some specific instances researchers can compare
different interfaces such as 2D and 3D against each other using the same
system, such as Sebrechts et al.~\cite{sebrechts:1999} study on visualization
of search results in text, 2D and 3D. Sebrecths et al.~\cite{sebrechts:1999}
et al.~approached the ``fairness dilemma'' by constructing the 2D interface by simply flattening their 3D
interface.

With regards to space time cube representation we believe that 
\emph{some} approach needs
to be taken to gain any clarity in the issue. However, unlike Sebrechts et
al.~\cite{sebrechts:1999}, we attempted to create a fair baseline given
assumptions on what we set out to investigate in our experiment, rather than
creating a baseline that is as closely related to the 3D system as possible. We focused on a baseline
comparison where both representations aim at providing users with an overall
understanding of the spatiotemporal patterns in the dataset at a glance. After
all, it is this advantage of space time cube representation that is most often
argued in the literature \cite{gatalsky:2004,kapler:2004,kraak:2003}. We rejected time sliders and animations
that partition the temporal dimension of datasets into discrete time steps,
because users cannot get an overview of the dataset at a glance with such
representations. Recognizing the limitation of granularity with ordinal
pseudocolor sequences \cite{ware:2004} we gave up any attempts of using sophisticated
color scales to reveal time information. Instead we compromised for an
approach where critical time points in 2D are indicated with semantic markup
(text), see Figure \ref{fig:2d}. This choice gives users the ability to perceive an
overview of the spatiotemporal patterns at a glance, even with 2D. Note that
the labels (markup) in Figure \ref{fig:2d} can easily be turned on or off with the
keyboard.

Both the space time cube system and 2D baseline system are interactive. With
the space time cube system users can for instance pan, rotate and zoom in and
out. With the baseline 2D system the user can toggle the display of time
labels and zoom in and out of a portion of the map.

It is important to note that the purpose of the baseline 2D system is to
provide the space time cube representation with a reasonable \emph{baseline}.
That is, the space time cube should ``beat'' the baseline in at least some
aspect to merit further research by the information visualization community.
The purpose of the 2D baseline system
is not to investigate how 2D visualization can be made more effective. This in
itself is an interesting research question, but out of scope of this paper.

\subsection{Space Time Cube System}
\begin{figure}
\begin{center}
\includegraphics[width=12cm]{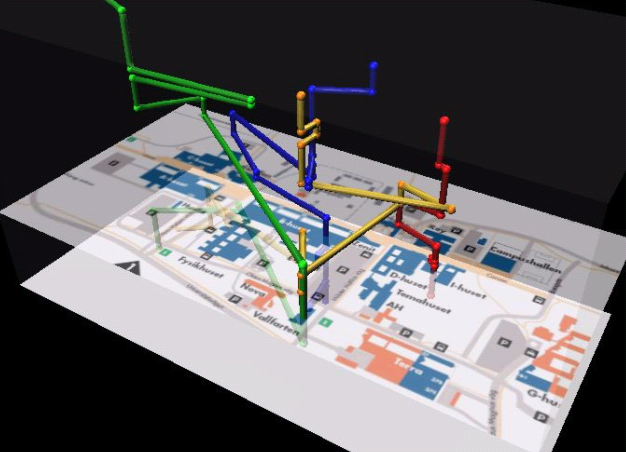}
\end{center}
\caption{Human walking data visualized in the space time cube system we
developed. Different colors represent different persons. When a person
stands still, the trace segment is perpendicular to the map plane.}
\label{fig:stc}
\end{figure}

To perform our investigation we developed a space time cube system capable of
rendering walking data traces inside a cube (see Figure \ref{fig:stc}). The system has a
``measurement'' plane that can be moved up and down along the height axis to
make it easier to read when a particular event occurred. The exact time of the
measurement plane's current position is displayed to the right of the space
time cube display area.

The space time cube system is controlled via either the keyboard or a
graphical user interface ({\sc gui}). Using the {\sc
gui} or a keyboard the user can rotate, zoom and move the measurement plane up
or down.

\subsection{Baseline 2D System}
\begin{figure}
\begin{center}
\includegraphics[width=12cm]{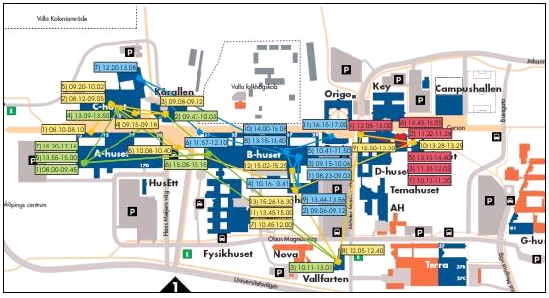}
\end{center}
\caption{Human walking data visualized in the baseline 2D system we developed.
Different colors represent different persons. In the figure the labels have
been turned on. Labels can be turned on and off with a press of a key.}
\label{fig:2d}
\end{figure}

The baseline 2D system displays walking data traces using different colors
(green, blue, yellow, red; see also Figure \ref{fig:2d}). The colors were the same as in
the space time cube system.

The colored line traces indicate different persons, and the labels indicate
the start and end time for a person at specific point in the map. Users can
toggle the display of labels and zoom in and out with the keyboard.

\section{Method}
We used a between-subjects experimental design where participants were exposed
to one of two conditions: either they used a space time cube system, or the
baseline 2D system.

Often within-subjects experimental design is preferable since 1) variation
between conditions is controlled within the participant; and 2) generally a
smaller number of participants are required. However, in this experiment it is
plausible that participants become increasingly familiar with the material
and task during the experiment. With a within-subjects design there is a risk that one
condition (call it condition A) better aids participants in understanding the material and the
task than the other condition (call it condition B).
This asymmetrical skill-transfer effect would in fact penalize the performance of condition A when
preceded by condition B, and unfairly benefit condition B when preceded by
condition A.
To avoid this confound a between-subjects design was used, and the number of
participants in the experiment was increased accordingly ($n = 30$).

\subsection{Participants}
30 participants, 15 male and 15 female, were recruited from the university
campus. The participants were screened for color blindness. None had any
previous experience in using information visualization tools. The groups were
gender-balanced.

\subsection{Apparatus}
The experiment was conducted on two laptops with 15$''$ screens and 32-bit color
depth. Although the physical dimensions of the laptop screens were identical,
the screen resolution varied slightly in the vertical dimension. The first laptop had a
screen resolution of 1280 $\times$ 1024 while the second laptop had a screen
resolution of 1280 $\times$ 800.

\subsection{Material}
To assess the participants’ understanding of the dataset, a set of 15
questions were designed. The questions were grouped into four different
question categories of varying difficulties and complexities according to
Andrienko et al.~\cite{andrienko:2003}. Along with a description of each question category we
supply an example from the material used in the conducted study (translated from Swedish).

\begin{description}
\item[Question Category 1] Simple ``when'' and simple ``what + where'': describes an
object's property at a given point in time, e.g. ``Where is the red person at
14:00?''

\item[Question Category 2] Simple ``when'' and general ``what + where'': describes the
situation at a given point in time, e.g. ``Are any two persons at the same place at
9:00?''

\item [Question Category 3] General ``when'' and simple ``what + where'': describes an
object's characteristics over time, e.g. ``Which buildings are visited by the
yellow person during the day?''

\item [Question Category 4] General ``when'' and general ``what + where'': describes
the development of an entire situation over time, e.g. ``Who is on the campus
area  for the longest time?''
\end{description}

15 questions were used in the experiment. Question categories 1--3 had 4
questions, question category 4 had 3 questions.

The questions were graded as either ``correct'' or ``incorrect'' based on a
predefined marking scheme.

\subsection{Procedure}
The participants were divided into two gender-balanced groups. One group used
the baseline 2D system while the second group used the space time cube system.
The experiment consisted of two sessions: a practice session and a testing
session. After the two sessions participants
were interviewed. The experiment was designed to require a maximum of one hour
of participants' time.

\subsubsection{Practice Session}
The first session was a practice session where participants were asked to
answer a set of written questions with the help of either system (space time
cube or the baseline 2D system). The practice session lasted around 20
minutes. The domain and the questions used for the practice session were
different from the material in the testing session. In the practice session
lightning strike data was used. The space time cube system visualized
lightning strikes as small red spheres in the cube. A corresponding system
generously provided by the Swedish Meteorological and Hydrological Institute
({\sc smhi}) was used as the practice baseline 2D system. The purpose of the
practice session was to introduce information visualization tools to the
participants and get them used to answering spatiotemporal questions with the
help of the system under investigation. The systems used in the practice
session were not designed to be directly comparable against each other.
Therefore we do not report the results from the practice session.

\subsubsection{Testing Session}
After a brief break participants proceeded with the testing session that
followed immediately after the practice session. The domain used in the
testing session was the human walking data, explained in Section 2
earlier. The procedure in the testing session was otherwise identical to the
one used in the practice session.

\section{Results}
Analysis of variance ({\sc anova}) was used for all statistical tests described in
this paper.

\subsection{Error}
The average error rate across all question categories for the baseline was 16\%
in comparison to 23\% for the space time cube representation. The difference is
not statistically significant (F$_{1, 28}$ = 4.167, p = 0.051), although very close
on the 0.05 level.

\begin{figure}
\begin{center}
\includegraphics[width=10cm]{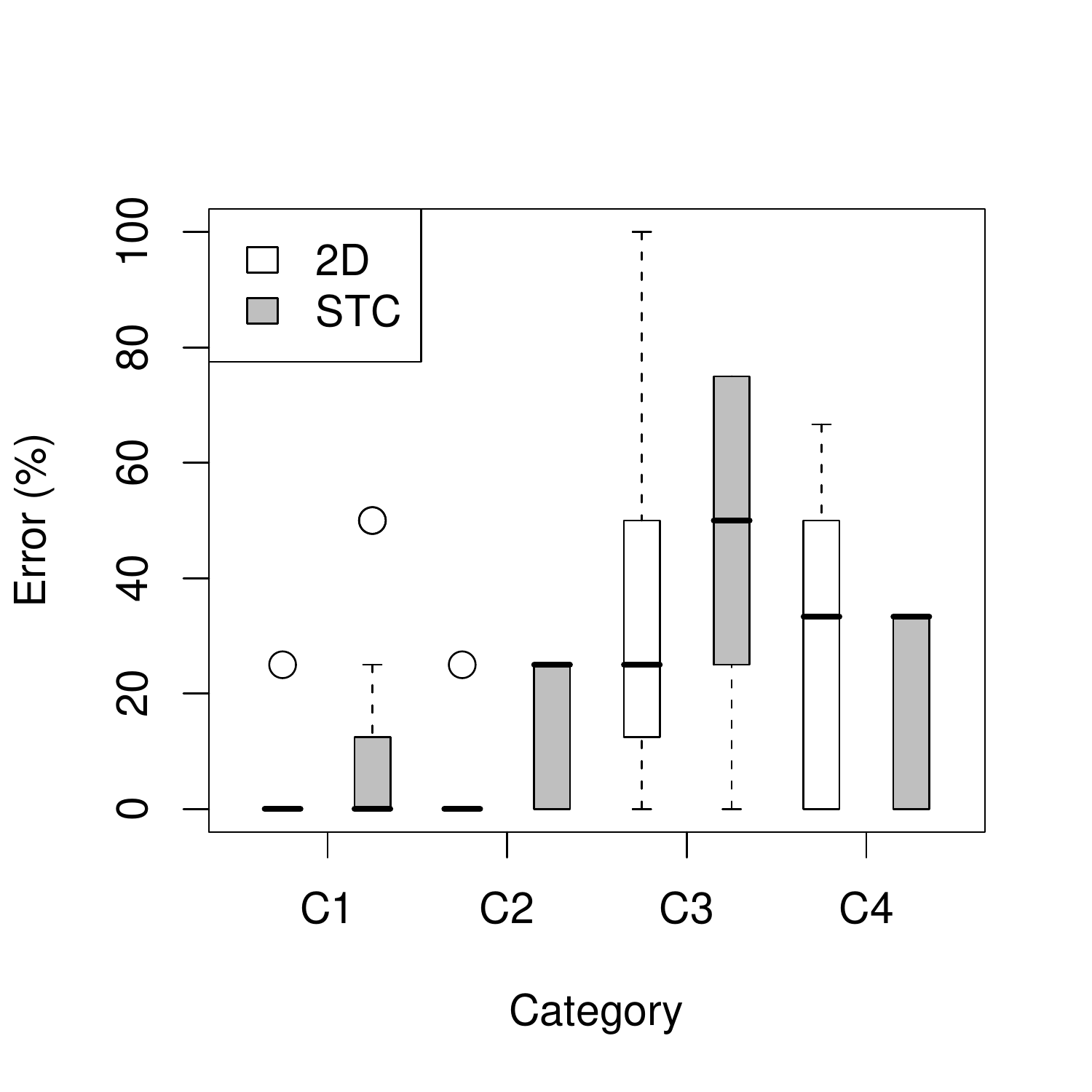}
\end{center}
\caption{Box-and-whisker plot of error rate (\%) as a function of
question category.}
\label{fig:err_cat}
\end{figure}

Breaking down error rates into individual question categories, error rates
were lower with the baseline 2D representation for the simple question
categories 1 and 2 that asked about objects' properties, or a situation, at a
given time. For these question categories the baseline 2D system had close to
0\% error rate (Figure \ref{fig:err_cat}). In question category 2 the baseline 2D system
resulted in significantly fewer errors (F$_{1, 28}$ = 9.800, p $<$ 0.005). Error rates
were particularly high in question category 3, but no statistical significant
difference between the systems was found (F$_{1, 28}$ = 2.343, p = 0.137). Question
category 4 was unique in the sense that the space time cube had a lower
average error rate in comparison to 2D (31\% for the 2D baseline system vs.
20\% for the space time cube). However,
the difference was not significant (F$_{1, 28}$ = 1.862, p = 0.183). From
the results it is clear that participants found it harder to answer
questions in categories 3 and 4 (cf. Figure \ref{fig:err_cat}).

These results are somewhat expected since the participants were novice users
of visualization tools, and only had a single session of practice before the
testing session. The fact that there was no statistical difference found
between the baseline 2D and space time cube system in neither question
category 3 nor category 4, shows that the higher error rates can most likely
be attributed to the difficulty increase of the question answering task in
general, rather than a particular deficiency in either system. Surprisingly, error
rates are more pronounced for question category 3 than category 4, even though
questions in category 4 demand much more understanding of the dataset than
questions in category 3.

\begin{figure}
\begin{center}
\subfloat{\includegraphics[width=6cm]{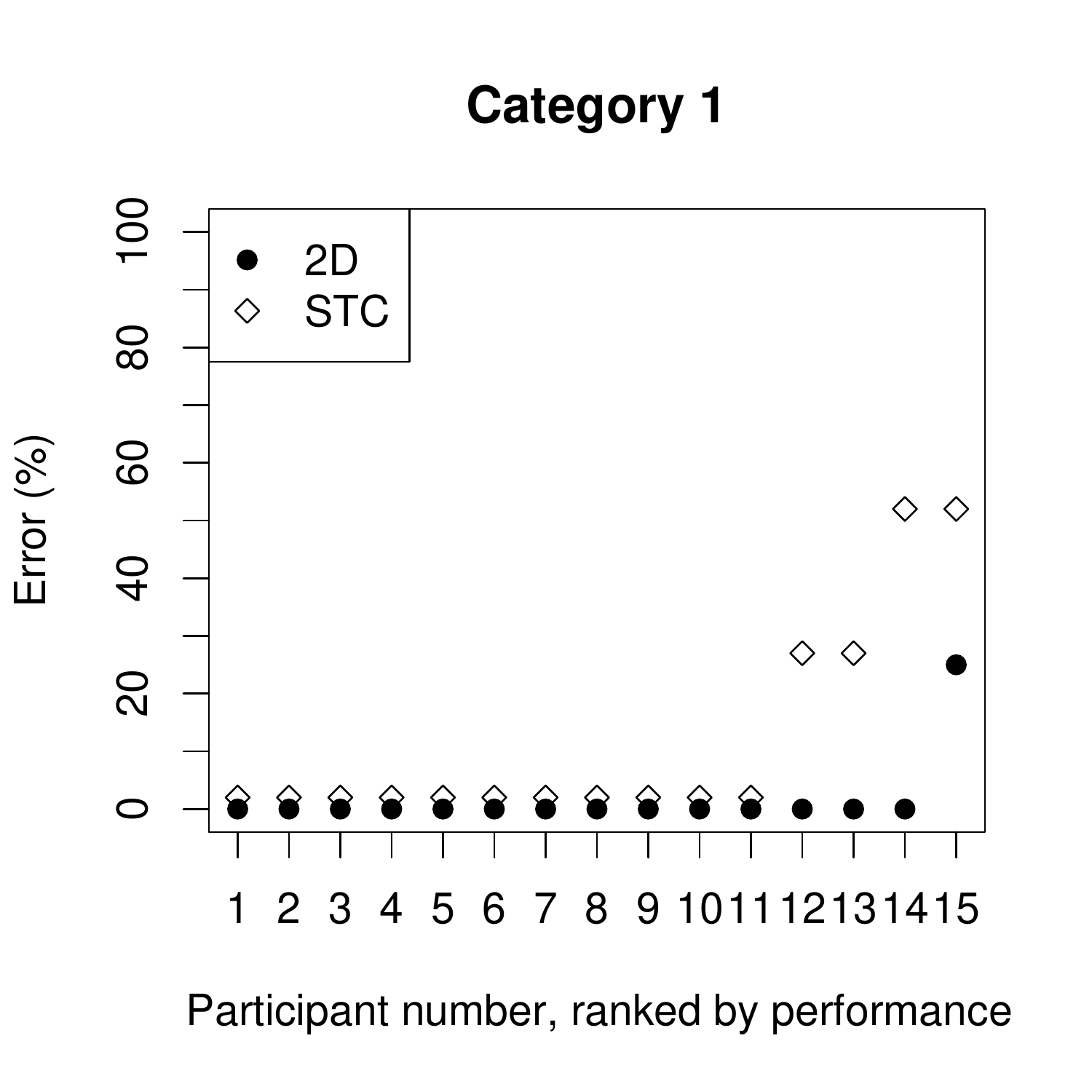}}
\subfloat{\includegraphics[width=6cm]{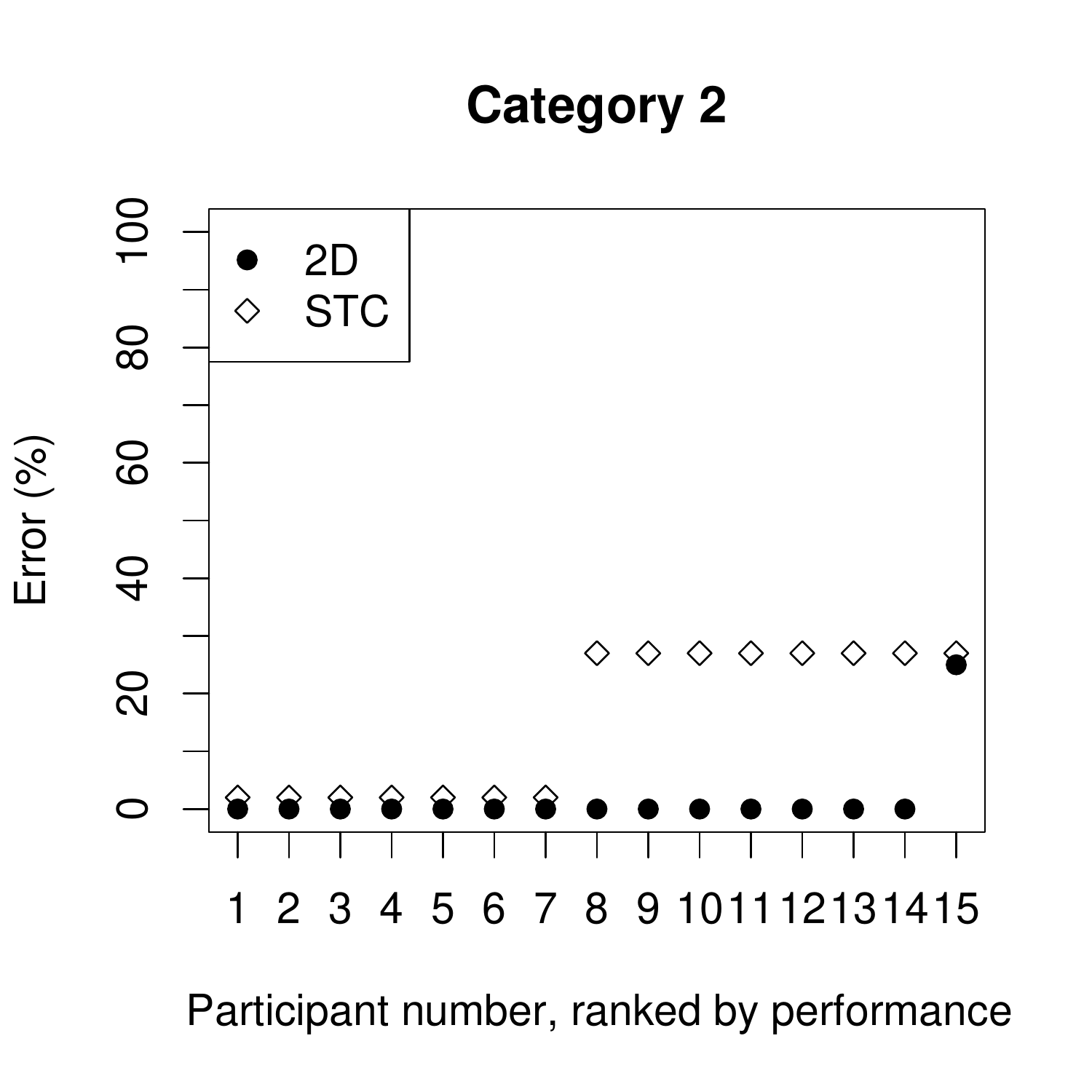}}\\
\subfloat{\includegraphics[width=6cm]{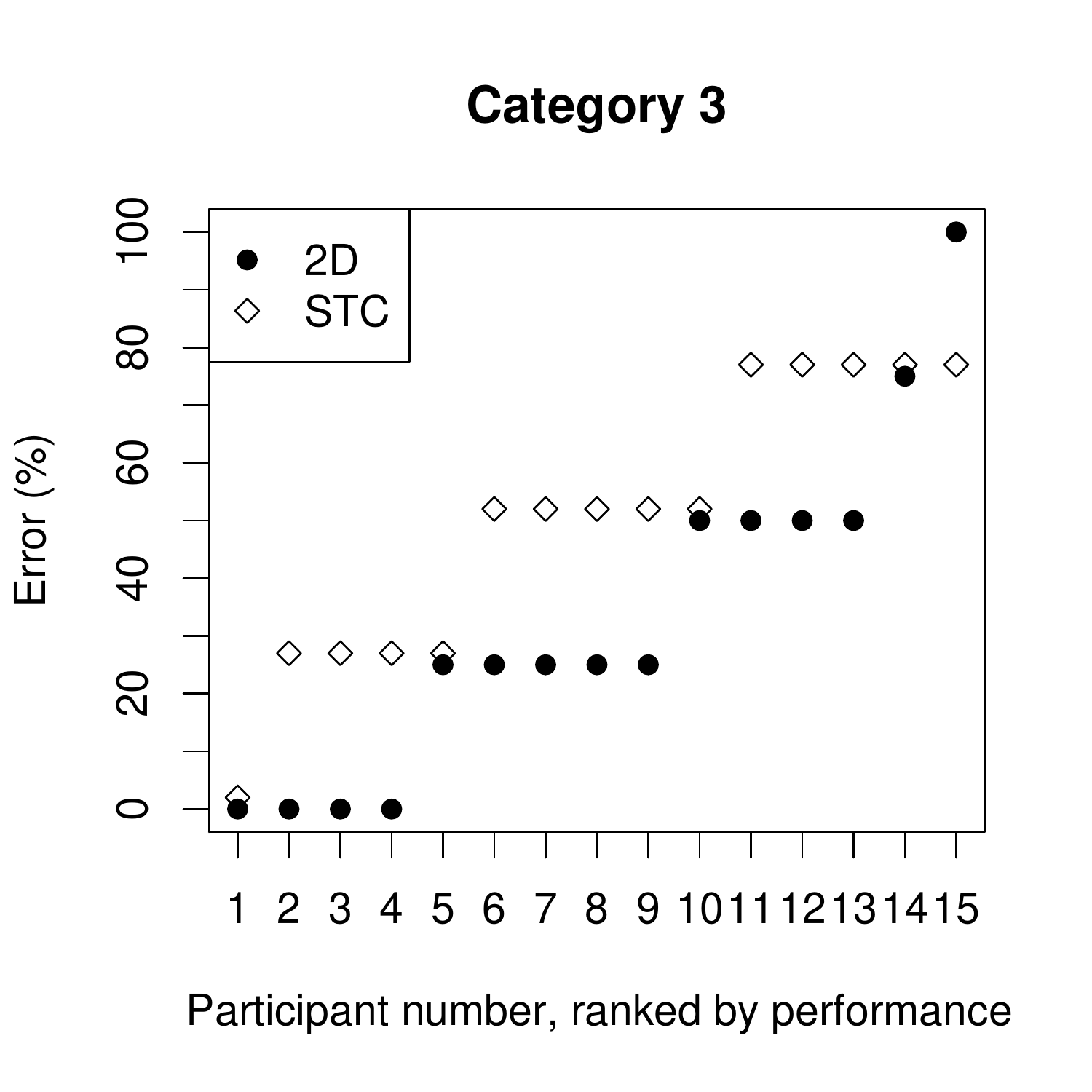}}
\subfloat{\includegraphics[width=6cm]{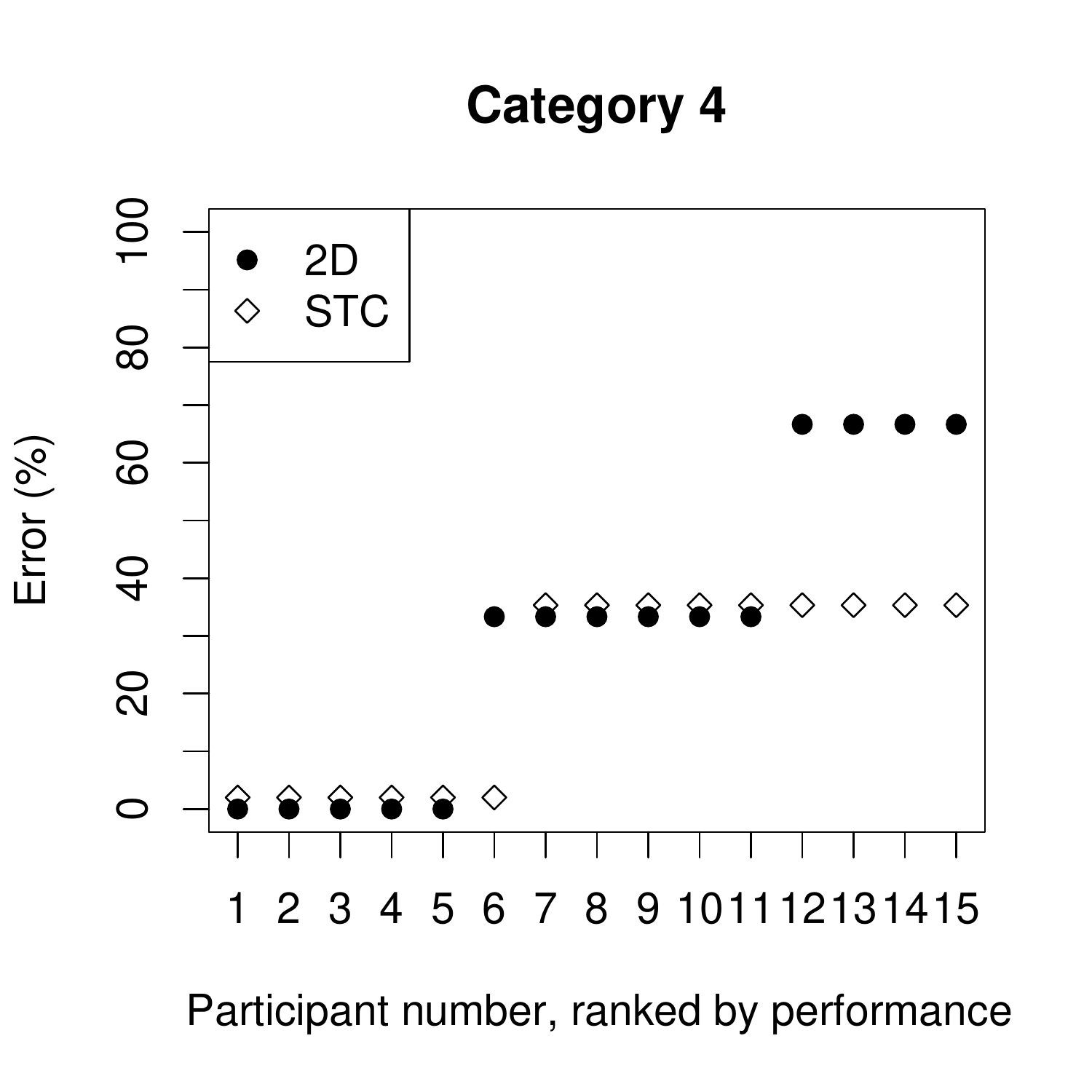}}
\end{center}
\caption{Error rate (\%) for question categories 1--4 as a function of
    participant.
The space time cube plot has been shifted slightly upwards graphically in relation
to the 2D baseline plot in order to make overlapping data points more easily
distinguishable. In the actual data, overlapping data points have identical
values.}
\label{fig:error_part}
\end{figure}

Figure \ref{fig:error_part} plots the error rate for individual participants in each condition
for question categories 1--4, ranked by performance (top performer using
baseline 2D representation against top performer using space time
cube representation, and so on). Question category 4 in Figure
\ref{fig:error_part} is particularly interesting because this question
category concerns the most difficult questions on the dataset. Note that, for question
category 4 in Figure \ref{fig:error_part}, at all
corresponding ranking positions every participant that used space time cube
representation consistently had the same or lower error rate than his or her
counterpart who used the baseline 2D representation.

\subsection{Response Time}
The average response time per question using the 2D baseline representation was 3
seconds lower (63 seconds) than space time cube (60 seconds). The result was
not significant (F$_{1, 28}$ = 0.217, p = 0.645).

\begin{figure}
\begin{center}
\includegraphics[width=10cm]{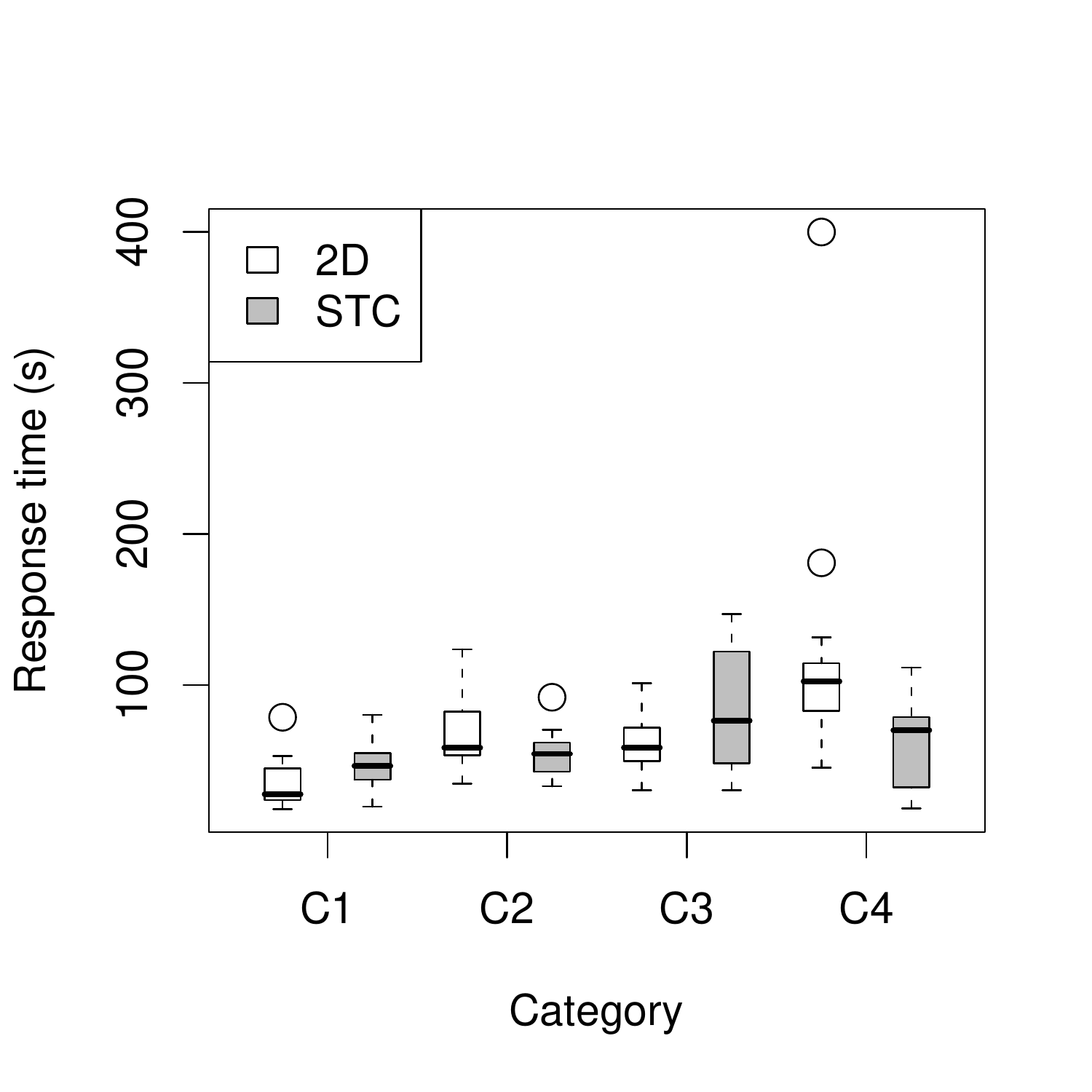}
\end{center}
\caption{Box-and-whisker plot of response time (s) as a function
of question category.}
\label{fig:rt_cat}
\end{figure}

Figure \ref{fig:rt_cat} shows the response times for the individual question categories. We
found a high-magnitude statistically significant difference in question
category 4 where space time cube representation halved the average response
time from 121 s in the baseline 2D system down to 60 s (F$_{1, 28}$ = 6.957, p
$<$ 0.05). This result supports the hypothesis that space time cube
representation is efficient in supporting users' understanding of complex
spatiotemporal patterns in datasets.

\begin{figure}
\begin{center}
\subfloat{\includegraphics[width=6cm]{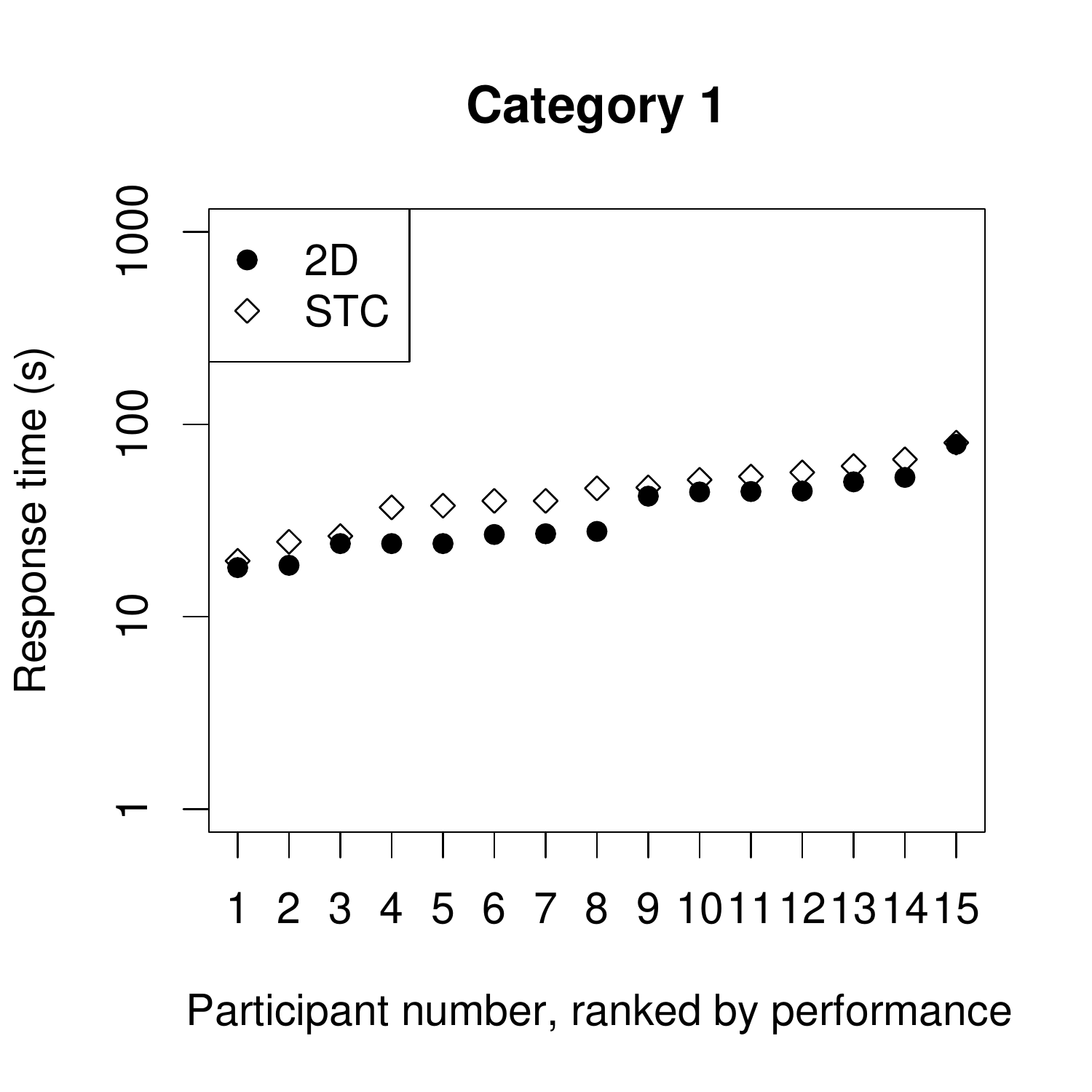}}
\subfloat{\includegraphics[width=6cm]{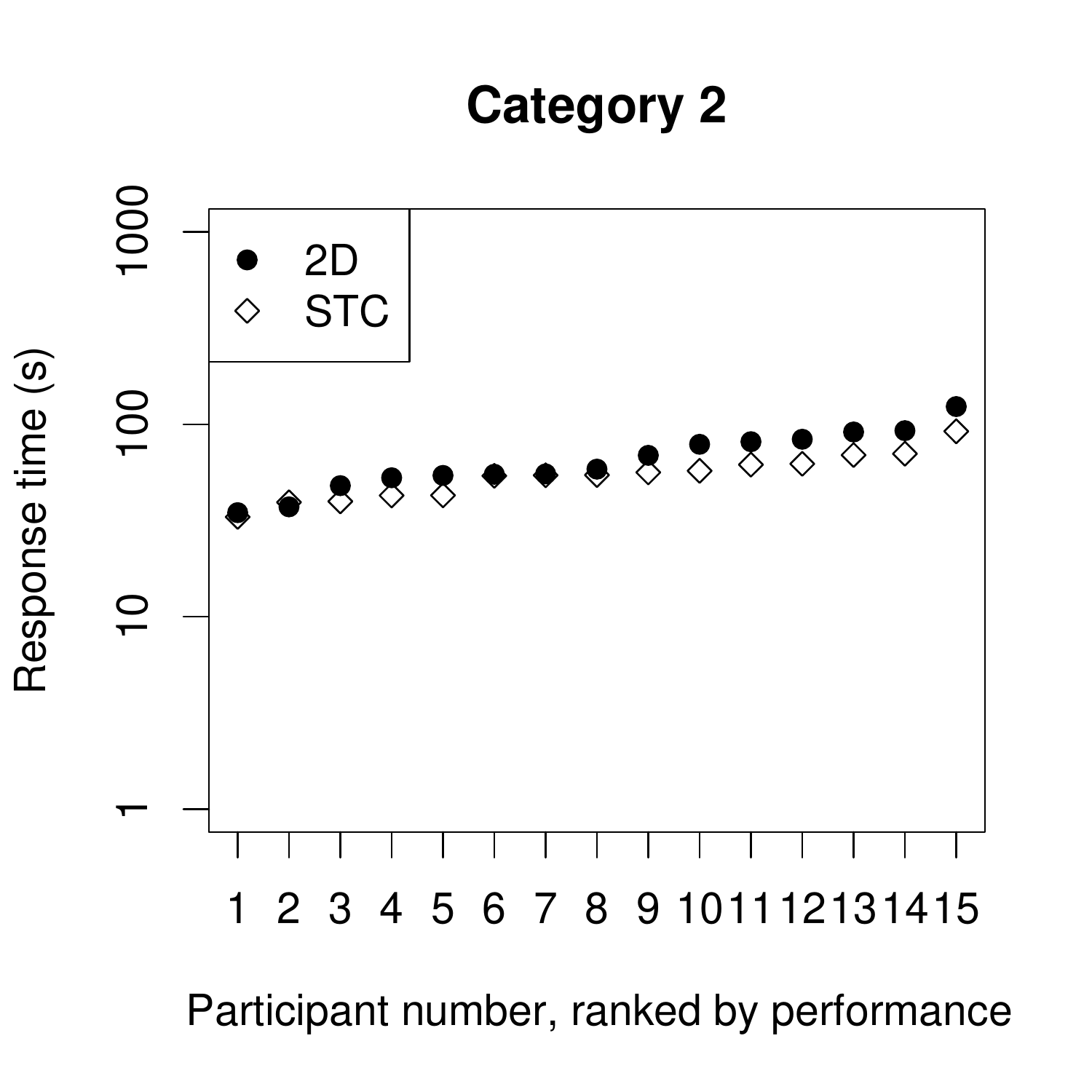}}\\
\subfloat{\includegraphics[width=6cm]{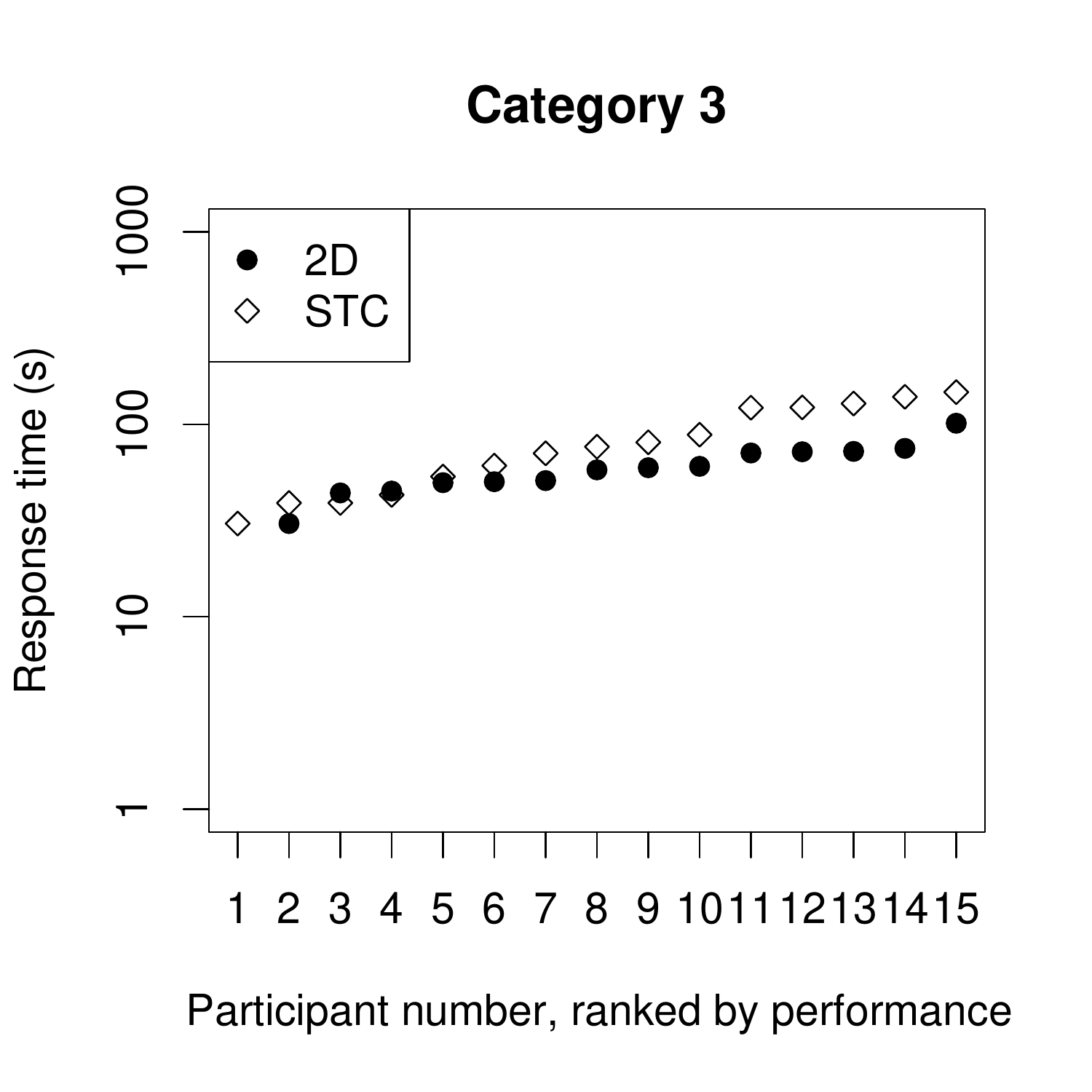}}
\subfloat{\includegraphics[width=6cm]{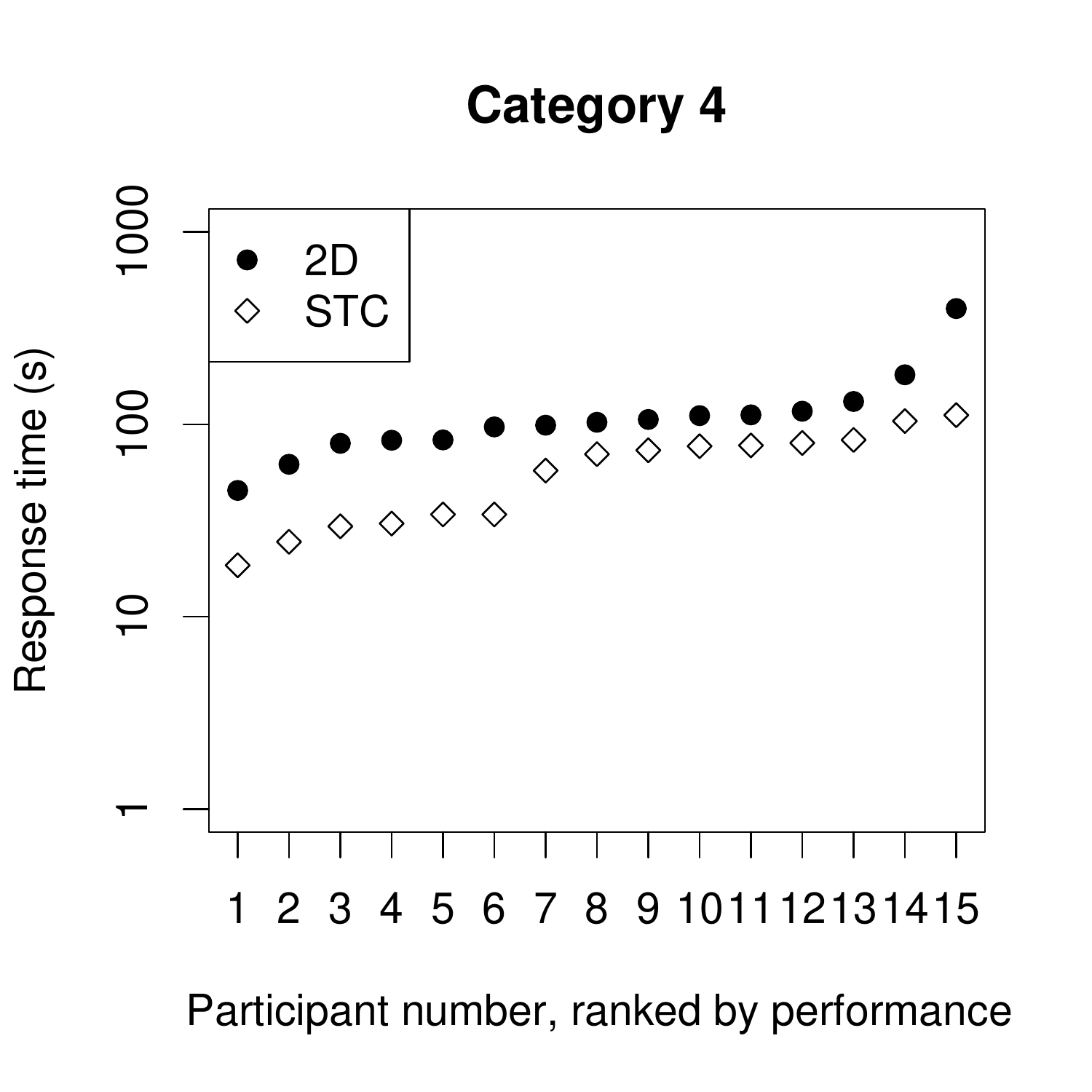}}
\end{center}
\caption{Average response time (s) for question categories 1--4 as
a function of participant number in each condition, ranked by performance.}
\label{fig:rt_cat4}
\end{figure}

Figure \ref{fig:rt_cat4} plots the response times for individual participants in each
condition for question categories 1--4, ranked by performance. As can be seen in
question category 4 in Figure \ref{fig:rt_cat4}, at all corresponding ranking positions every participant using space
time cube representation consistently outperformed his or her counterpart
using baseline 2D representation.

\subsection{Open Comments}
Participants gave us some open comments at the interview part in the experiment. When
interpreting these comments it is important to keep in mind that participants
had only experienced one representation.

The baseline 2D system was perceived as easy, interesting to use, ``fun'' and
participants thought it had a ``professional'' feel. Eight participants stated
that they thought the interconnected lines made the visualization easier to
interpret, one participant stated the opposite.

Space time cube representation was perceived as intuitive, engaging, easy to
understand and ``cool''. Three participants stated difficulties with using the
measurement plane (along the time axis). Eight participants explicitly stated
that they had no problem manipulating the measurement plane.

\section{Conclusions}
In relation to the research questions we posed in the introduction of this
paper, we found that novice users could indeed work effectively with the space
time cube representation after a short amount of practice. Overall there are
no measurable performance differences in either error rates or response times
between the space time cube system and the baseline 2D system. However, in
individual question categories significant differences in both error rates and
response times were found. It has been argued that the real benefit of the
space time cube is in supporting users when observing nontrivial
spatiotemporal patterns that require a ``bird's-eye'' view of the dataset
\cite{gatalsky:2004,kapler:2004,kraak:2003}. The dramatic reduction in response times for the most complex and
demanding questions in category 4 supports this hypothesis.

Our results also show that novice users are generally more error prone when
answering a category of ``simple'' questions, such as ``Are any two persons at the
same place at 9:00?'' (question category 2), when using space time cube
representation. When developing systems that are expected to be used by
non-experts (e.g. teaching support), we suggest implementing an alternative
visualization view that more effectively aids novice users’ perception of
individual data points at specific locations or points in time.

It is hard and perhaps misleading to attempt to directly generalize the implications of the results
found in any experiment that compares two information visualization systems against each
other. Clearly different material (both questions and 
map choice) might affect the experimental outcome. Nevertheless, it is
worth emphasizing that every effort was made to ensure that the experiment would be
as unbiased as possible. For example, the map and the walking data
participants analyzed was real data and not artificially constructed for the
purpose of the experiment. Moreover, the questions asked were designed
and distributed into several categories according to the formalism proposed by
Andrienko et al.~\cite{andrienko:2003}. It is perhaps best to view our
contribution as supporting a hypothesis, call it the ``space time cube
hypothesis''. Essentially this hypothesis says that space time cube
visualization aids users in analyzing complex spatiotemporal patterns. In
general, one cannot prove a hypothesis, only disprove it. We did not disprove
the ``space time cube hypothesis''---in fact our results show that users
are significantly faster with a space time cube representation when answering
questions on complex spatiotemporal patterns.
Our final conclusions are twofold. First, everything else equal, for complex spatiotemporal patterns
there is no reason to believe space time cube representation would not result in
faster response times than a baseline 2D representation similar to the one used
in our experiment. Second, space time cube is worth further consideration and
investigation by information visualization researchers. Prior to
the experimental results presented here, there was no hard empirical
motivation that space time cube representation had any benefit \emph{at all}.
From our results we can deduce that  space time cube is at least worth further
investigation, for example by varying data density, choice of maps, domains or level of expertise
among the participants.

\section{Acknowledgments}
This work was in part sponsored by {\sc sics} Link\"oping. We thank Piotr
Zieli\'{n}ski and David Stern for their assistance.

\end{document}